\documentclass[prl,aps,reprint,twocolumn,nofootinbib,tightenlines,nobibnotes,notitlepage,preprintnumbers,superscriptaddress,nobalancelastpage]{revtex4}
\pdfoutput=1
\usepackage{wrapfig}
\usepackage{verbatim}
\usepackage{amsmath, multirow, amssymb, wasysym, gensymb}
\usepackage{graphicx}
\usepackage{amsfonts}
\usepackage{graphics}
\usepackage[ngerman, english]{babel}
\usepackage{float}
\usepackage{subfigure}
\usepackage[utf8]{inputenc}
\usepackage[usenames,dvipsnames]{xcolor}
\usepackage[normalem]{ulem}
\usepackage{relsize}
\usepackage{slashed}

\newcommand{\gsim}{\raise0.3ex\hbox{$\;>$\kern-0.75em\raise-1.1ex\hbox{$\sim\;$}}} 
\newcommand{\lsim}{\raise0.3ex\hbox{$<$\kern-0.75em\raise-1.1ex\hbox{$\sim\;$}}}

\usepackage{subfigure}

\input{tcilatex}
\begin{document}

\title{Testing keV sterile neutrino dark matter in future direct detection experiments}
\author{Miguel D.\ Campos}\email{miguel.campos@mpi-hd.mpg.de}
\affiliation{Max-Planck-Institut f\"ur Kernphysik, Saupfercheckweg 1, 69117 Heidelberg,
Germany}
\author{Werner Rodejohann}\email{werner.rodejohann@mpi-hd.mpg.de}
\affiliation{Max-Planck-Institut f\"ur Kernphysik, Saupfercheckweg 1, 69117 Heidelberg,
Germany}

\begin{abstract}
\noindent
We determine constraints on sterile neutrino warm dark matter through direct detection 
experiments, taking XENON100, XENON1T and DARWIN as examples. 
If keV-scale sterile neutrinos scatter inelastically with bound electrons of the 
target material, an electron recoil signal is generated. This can be used to set 
limits on the sterile neutrino mass and its mixing with the active sector. While not 
competitive with astrophysical 
constraints from $X$-ray data, the constraints are the first direct laboratory bounds on 
sterile neutrino warm dark matter, and will be in some parts of parameter space 
the strongest limits on keV-scale neutrinos.   
\end{abstract}

\maketitle

\date{\today }

\section{Introduction}
\label{s:intro}
Dark Matter (DM) contributes five times more to the energy budget of the 
Universe than ordinary matter \cite{Ade:2015xua}. 
While first hints of the existence of DM 
emerged more than 80 years ago \cite{Zwicky:1933gu}, its nature has remained 
elusive, and despite intense effort in the scientific community 
(both theoretical and experimental) 
scarce progress has been achieved to settle the issue. 
Since modifications of gravity cannot explain the evidence for 
DM on all length and time scales, the generally accepted approach to the 
DM problem is that particle physics lies at its origin. 
There is however no lack of candidates, the most popular 
ones belonging in the class of Cold Dark Matter (CDM), whose 
properties are in convincing agreement with structure formation 
at large scales \cite{Springel:2005nw}. 
The most frequently considered CDM candidates are Weakly Interacting 
Massive Particles (WIMPs), which are motivated 
by many theories beyond the Standard Model of Particle Physics, and 
can generate the right abundance of DM in a straightforward way. Indeed, a large 
number of dedicated direct detection experiments 
\cite{Undagoitia:2015gya} 
search for the nuclear recoil from WIMP-nucleon scattering. 
Apart from such direct detection, indirect detection is possible by
looking for signals of decay or annihilation of DM particles 
in astrophysical observations. The third way of looking for cold dark
matter particles is to produce them in collider experiments. 
A recent review of the various DM candidates, search principles, and the 
current experimental situation can be found in Ref.\ \cite{Klasen:2015uma}.

However, some tension has arisen in the last decade with observations 
at small galaxy formation scales known as 
the cusp-core problem \cite{Wyse:2007zw}, 
the missing satellite problem \cite{Moore:1999nt} and 
the too-big-to-fail problem \cite{BoylanKolchin:2011de}. 
This has motivated many studies on Warm Dark Matter (WDM) particles, 
whose properties lead to similar structure formation on 
large scales, but due to a longer free-streaming length to 
less structure at smaller ($\lsim 0.1$ MPc) 
scales \cite{Boyarsky:2008xj}.  
The most prominent WDM candidates are sterile neutrinos with masses 
of several keV and very small mixing with active neutrinos. Such 
sterile neutrinos might, thus, be connected to the mass generation of 
light active neutrinos \cite{Canetti:2012kh,Abazajian:2012ys}. 
Various possibilities for the generation of the WDM density are 
possible, e.g.\ via vacuum oscillations \cite{Dodelson:1993je}, 
or resonant oscillations with the help 
of lepton asymmetries \cite{Shi:1998km}, see \cite{Kusenko:2013saa} 
for more discussion. We will not specify any production mechanism here. 
Indirect searches are possible as their electroweak decay generates 
characteristic $X$-ray 
signals (it is worth mentioning that hints for an unidentified 
line at about 3.5 keV \cite{Bulbul:2014sua,Boyarsky:2014jta} are currently 
under active 
discussion and investigation). This approach is the analogue of the
indirect detection method for WIMPs mentioned above.   
The production of keV neutrinos in $\beta$-decays 
is another possibility \cite{Mertens:2014nha} to search for those 
WDM candidates, this is the analogue of WIMP collider searches.\\
In this paper we will show that obtaining 
{\it direct detection limits} on keV WDM sterile neutrinos is possible in 
ex\-pe\-ri\-ments originally aimed at direct CDM or WIMP detection: 
if the sterile neutrino mixes with active neutrinos and, thus, 
scatters inelastically with bound electrons of the target material, the 
scattered electrons create electron recoils which some of the direct 
detection experiments can detect. We use data from the 
XENON100 experiment to set the first direct detection 
limits on keV neutrino WDM, and present the prospects for the upcoming 
 XENON1T, XENONnT and DARWIN experiments. Those bounds are not compatible with indirect 
astrophysical bounds, but will be the only direct detection limits, and 
in some parts of parameter space the best limits on sterile neutrinos, 
assuming, of course, that they form the wholeness of dark matter and mix with 
electron neutrinos.

\section{Direct detection of dark matter}
\label{s:dddm}
\begin{figure}[b]
\centering
\subfigure{
\includegraphics[height=0.2\textheight]{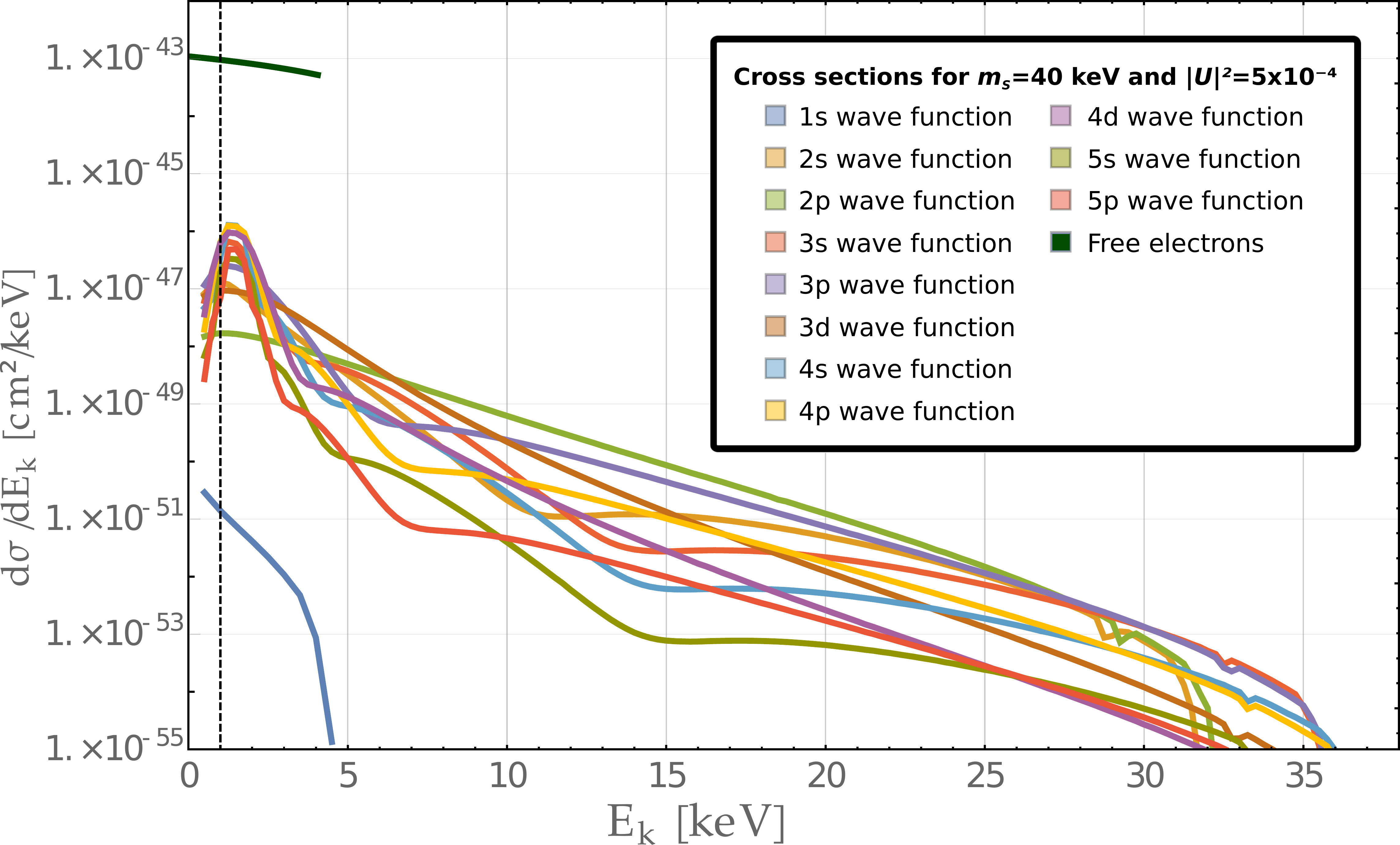}}
\caption{\label{fig:ds} 
Differential cross section of massive sterile neutrinos with free (dark green) and bound electrons (light colors) for $m_S=40$ keV and $|U_{Se}|^2=5\times10^{-4}$. Dashed vertical line represents lower threshold of $E_{\rm Th}=1$ keV (see next section).
}
\end{figure}

Direct detection experiments are suited for WIMP physics through their 
coherent scattering with a nucleus, and observing the  
generated nuclear recoil. 
Constraints are set on interaction cross section and DM mass, and 
currently the direct detection experiments setting the most stringent 
limits for masses above $\sim 5$ GeV are XENON100 \cite{Aprile:2016swn}, LUX \cite{Akerib:2016vxi} and Panda-X \cite{Tan:2016zwf} involving several dozen kilograms of liquid xenon. We will focus in what follows on the
former experiment, whose next step is called XENON1T \cite{Aprile:2015uzo}, involving 
1 ton fiducial volume; an even bigger stage would be XENONnT. 
We will also estimate the effects in the future DARWIN experiment \cite{Aalbers:2016jon}, that will aim at a 100-fold increase in exposure (and sensitivity to spin-independent WIMP-nucleus interaction) compared to XENON1T.
The detection principle (for a general overview see \cite{Undagoitia:2015gya}) 
of these dual phase detectors is that the
de-excitation of secondary xenon atoms that have interacted with the primary scattered atom lead to a prompt
scintillation signal (called S1). Ionization electrons are
extracted in the gas phase part of the instrument via electric fields and 
create another light signal (S2) via scintillation.  
For WIMP searches, electron recoils (ERs) represent a source of
background. They can be produced by the scattering of photons or electrons with
the electron cloud surrounding the xenon nucleus, and the sources 
include intrinsic $\beta$ decays of element traces present in 
xenon (such as $^{222}$Rn and $^{85}$Kr) and
gamma rays crossing the shielding. 
As ERs and nuclear recoils have different ratios of S1 and S2, it
is possible to impose cuts to exclude ERs. However, when one thinks
beyond the WIMP possibility, ERs are interesting on their own and 
allow us to probe physics beyond standard WIMPs. Indeed, dedicated
analyses of the XENON100 collaboration have studied axions \cite{Aprile:2014eoa}
and 
DM particles interacting purely with electrons \cite{Aprile:2015ade}.  
We use here the results from \cite{Aprile:2014eoa} to set constraints on
keV-scale sterile neutrino dark matter, and will also give limits on
the relevant parameters for future stages of the 
experiment\footnote{Let us note that other current or future experiments with sensitivity for electron recoils such as CDMS, CRESST, EDELWEISS, LZ or XMASS can also provide interesting limits.}.
To be able to use the ER data coming from direct detection experiments it is necessary that the signal produced by the DM candidate exceeds the background in some energy interval. As we will see, it is possible to generate the required excess when we consider WDM sterile neutrinos in the keV range. We use the background model coming from calibration data for ER in XENON100 \cite{Aprile:2014eoa}. 
For the case of XENON1T we will scale this background model such that it reproduces the background of $\sim 1.8\times 10^{-4}$ (kg day keV)$^{-1}$ predicted in \cite{Aprile:2015uzo} for XENON1T. We will also use the predictions from \cite{Schumann:2015cpa} and \cite{Aalbers:2016jon} to estimate the effect in the DARWIN experiment.
As a comparison the average background in DARWIN is estimated to be $\sim 2.05\times 10^{-5}$ (kg day keV)$^{-1}$, using an extrapolation of Fig. 2 (right) in \cite{Schumann:2015cpa}.\\
The process analyzed is based on the inelastic charged and neutral 
scattering $N_S \, e^-\to\nu_e e^-$ (and $N_S \, e^-\to\bar\nu_e e^-$),  
where a WDM sterile 
neutrino $N_S$ 
mixing with an 
active state scatters an electron from a xenon atom. 
Its velocity in the standard halo model 
is $v(N_S) \simeq 220$ km/s ($\beta=\mathcal{O}(10^{-3})$) and hence its energy essentially 
equals its mass, $E_S \simeq m_S$. 
The cross section for the process in the case of free electrons at rest (i.e.\ $p_e=(m_e,\vec{0})$ for the electron at rest
 and $p_e^\prime=(m_e+E_k,\vec{p}_e)$ for the scattered one) is
\begin{equation}\label{eq:dsfree}
\begin{aligned}
\dfrac{d\sigma_{\text{free}}}{dE_k} = & 2\dfrac{G_F^2}{\pi}|U_{Se}|^2\dfrac{m_e}{|\vec{p}_{S}|^2} \left[g_1^2E_S\left(E_S+\dfrac{m_S^2}{2m_e}\right)\right.  \\
& + g_2^2(E_S-E_k)\left(E_S-E_k+\dfrac{m_S^2}{2m_e}\right)\\
& - \left. g_1g_2(m_eE_k+\tfrac{1}{2}m_S^2)\right]. 
\end{aligned}
\end{equation}
The total cross section is the sum of the $\nu_e e^-$ and $\bar{\nu}_e e^-$ final states and is 
dominated by the $\nu_e e^-$ channel. We have defined  
\begin{equation}
\nonumber
g_1^{\nu}=g_2^{\bar{\nu}}:=1+ \frac 12 (g_V+g_A)\,,~  
g_2^{\nu}=g_1^{\bar{\nu}}:=\frac 12 (g_V-g_A)\,,
\end{equation}
with $g_V = -\frac 12 + 2 \sin^2 \theta_W $ and $g_A = -\frac 12$. 
The reaction, as in all sterile neutrino processes, is suppressed due to the 
mixing angle between sterile and active neutrinos $|U_{S\alpha}|^2$. 
We note that Ref.\ \citep{Ando:2010ye} has considered coherent 
inelastic WDM-atom
scattering of a 5 keV sterile neutrino, not taking the effect of 
bound electrons into account.  
In contrast, we will consider here sterile neutrinos in the range 
$\mathcal{O}(10-50)$ keV, corresponding to a wavelength of 
$\mathcal{O}(10^{-8}-10^{-9})$ cm. As the radius of a xenon atom 
is $\sim1.1\times10^{-8}$ cm, the electron-neutrino scattering is
incoherent and all the bound electrons in the xenon atom must be
considered. As scattering with bound electrons leads to larger
recoil than the free case (see Fig.\ \ref{fig:ds}), this is actually crucial to our analysis.  
One of the reasons this is important is because when considering just free electrons one would need masses 
higher than $\sim\,$20 keV to go beyond the minimum threshold of the detector, discussed later, hence entering  the 
incoherent regime.
To evaluate the electron recoil cross section with these bound electrons, we have performed the calculation 
in a similar way as it was done in Ref.\ \cite{Gounaris:2004ji},
suitably modified to take the non-negligible incoming neutrino
mass into account. An effective mass for the bound electron is defined as 
$\tilde m := E_B^2-|\vec{p}_B|^2\equiv E_B^2 - p_B^2$, 
where the bound electron has momentum ${\vec{p}_B}$ and energy $E_B=m_e-\varepsilon$, $\varepsilon$ being the binding energy.
All atomic wave functions have been considered using the Roothann-Hartree-Fock method, with data coming from Ref.\ \cite{bunge1993roothaan}. It is worth to mention that neither electron spin nor relativistic effects were taken into account in this expansion. It is suggested in \cite{Gounaris:2004ji} that errors for the most loosely bound electronic states could be up to $10\%$ due to spin effects.
The differential cross section for an electron in a state $t$ ($t=1s$, $2s$, $2p$,$\ldots$), in the rest frame of the atom where $(p_B,\theta,\phi)$ are the variables of the bound electron, is obtained through
\begin{equation}
\label{dst}
\begin{array}{l}
\dfrac{d\sigma_t}{dE_k} = \int \dfrac{p_B^2dp_Bd(\cos\theta)
  d\phi}{(2\pi)^3}\dfrac{|R_t(\vec{p}_B)|^2}{4\pi} \\ 
\dfrac{|\mathcal{M}|^2}{4E_SE_B|\beta-p_B/\tilde
  m|}\dfrac{1}{8\pi \lambda^{1/2}(s,m_S^2,\tilde
  m^2)}\left|\dfrac{du}{dE_k}\right|. 
\end{array} 
\end{equation}
Here, $R_t(\vec{p}_B)$ are the radial wave functions of the bound electrons defined in \cite{Gounaris:2004ji} and normalized such that
\begin{equation}
\int_0^\infty \dfrac{k^2dk}{(2\pi)^3}|R_t(k)|^2=1\,.
\end{equation}
The function $\lambda(a,b,c):=a^2+b^2+c^2-2ab-2bc-2ca$ is the K\"allén function and $s$ and $u$ are the usual Mandelstam variables.
As an example, the differential cross section for free and bound electrons in the different shells is presented in Fig.\ \ref{fig:ds} for a given choice of parameters, namely $m_S=40$ keV and $|U_{Se}|^2=5\times10^{-4}$. 
As we can see in Eq.\ \eqref{dst}, the differential cross section
involves an integration in $p_B$. The upper limit for this variable
is obtained from the allowed range of the final state particles as it was done in \citep{Liao:2013jwa}, which contains the threshold condition for the process to occur in the different bound states (i.e.\ $E_S>\varepsilon_t$). The limits on $E_k$ were obtained from \cite{Gounaris:2004ji} but they are equivalent to those presented in \citep{Liao:2013jwa}. These limits are dependent on $\varepsilon_t$, a fact reflected in Fig.\ \ref{fig:ds} where the most tightly bound electron ($\varepsilon_{1s}=34.561$ keV) has a differential cross section with a narrow allowed range in $E_k$ in strong contrast to the more loosely bound states ($\varepsilon_{2s}=5.453$ keV, $\varepsilon_{2p}=4.893$ keV,$\ldots$).

We can obtain the differential event rate of the process, in units of (kg day keV)$^{-1}$
\begin{equation}
\dfrac{dR_t}{dE_k}(m_S,|U_{Se}|^2) = \dfrac{\rho_0}{m_S}n_e\int \dfrac{d\sigma_t}{dE_k}(m_S,|U_{Se}|^2) f(\beta)\beta d\beta\,.
\end{equation}
Here, $\rho_0$ is the local DM density, $n_e$ the number of electrons per kilogram of target material and $f(\beta)$ is the velocity distribution in the detector frame.
In general, different DM production mechanisms lead to different distribution functions; however, one can safely assume that this distribution has evolved with time, leading to a DM halo with a (truncated in $v_{\rm esc}=544$ km/s) Maxwell-Boltzmann distribution, centered in $v_c=220$ km/s as it is done in the standard halo model\footnote{As the analysis follows the usual WIMP strategy, 
effects such as annual modulation are, in principle, also observable but are neglected here.}. 

If $T$ is the exposure time and $M$ the mass of the detector we can finally define the differential number of events corresponding to a particular electronic configuration:
\begin{equation}\label{eq:NE}
\dfrac{dN_t}{dE_k}(m_S,|U_{Se}|^2) = M \cdot T \cdot \dfrac{dR_t}{dE_k}(m_S,|U_{Se}|^2)\,.
\end{equation}
For XENON100 and XENON1T we take respectively:
\begin{equation}
\label{MT}
\begin{aligned}
M_{100}=34 \quad\text{kg} & \quad\text{and}\quad T_{100}=224.6 \quad\text{days}\\
M_{\rm 1T}=10^3 \quad\text{kg} & \quad\text{and}\quad T_{\rm 1T}=2\times 365 \quad\text{days}.
\end{aligned}
\end{equation}
Notice that in the case of XENONnT a global factor n appears in comparison with XENON1T. 
Several experimental aspects will probably be slightly different from the case of XENON1T, 
and cannot be reliably stated at the moment. The analysis for XENONnT has not been 
included here due to these reasons, but the 
limits can roughly be estimated to scale with n. For the DARWIN case we will use an exposure of $M\cdot T= 200$ ton$\times$year as in \cite{Aalbers:2016jon}.

\section{Analysis and results}
\label{s:results}
\begin{figure}[t]
\centering
\includegraphics[width=0.4\textwidth,height=4.25cm]{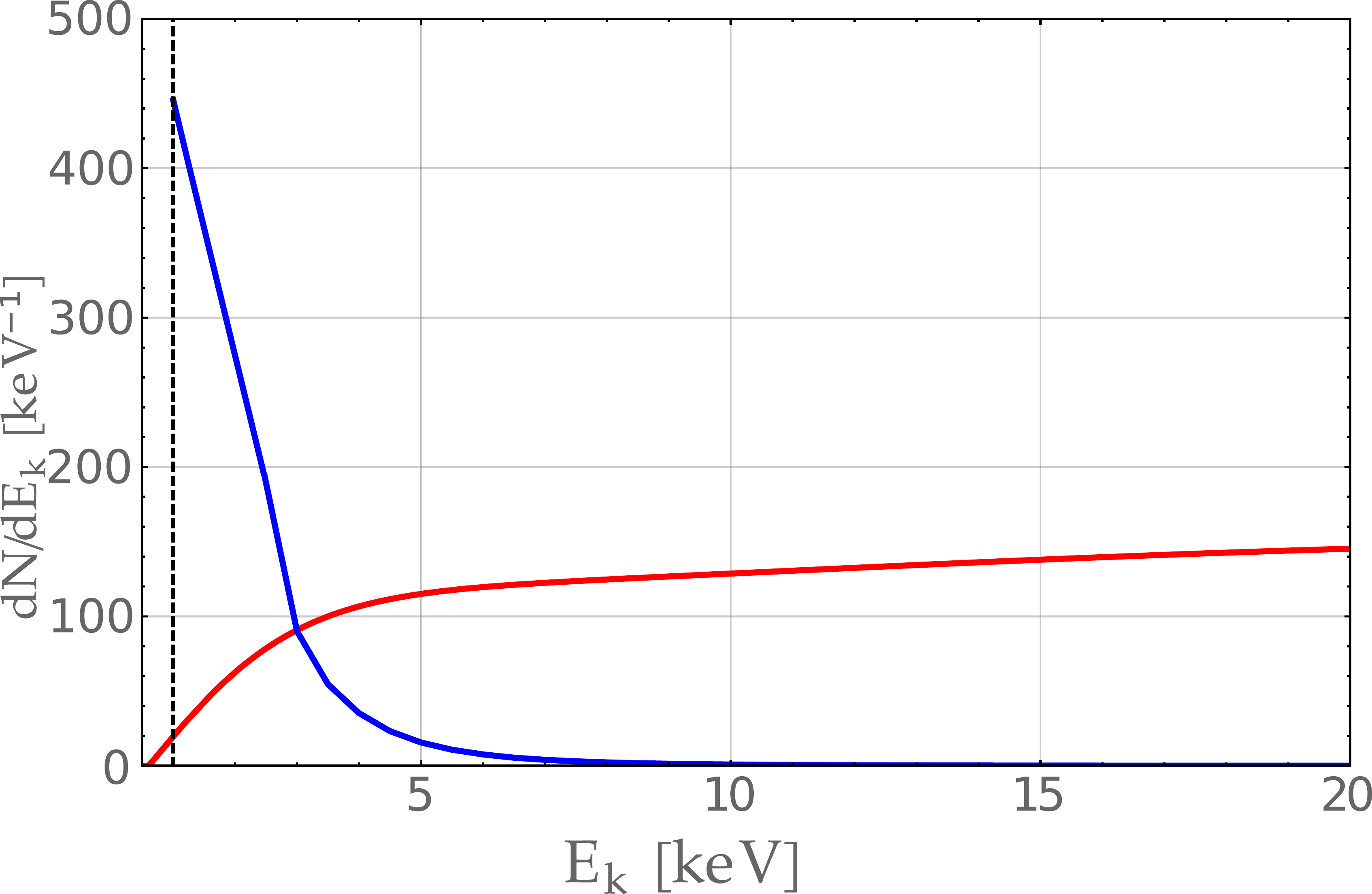}
\caption{Differential number of events for bound electrons for $m_S=40$ keV and $|U_{Se}|^2=5\times10^{-4}$ in XENON1T (blue) and estimated background $F_b$ (red). Dashed vertical line represents lower threshold of $E_{\rm Th}=1$ keV.}
\label{fig:dN}
\end{figure}

We perform now an analysis to estimate the parameter range for the variables $(m_S,|U_{Se}|^2)$ that can be excluded 
using a direct detection experiment. To be able to evaluate the differential number of events and compare it with the background it is necessary to take into consideration the global acceptance of the detector for ER, $\rm Acc(PE)$, which is specially important in the low energy range and depends on the number of photoelectrons ($\rm PEs$) measured in S1. Equally important is the conversion function ${\rm Conv}(E_k)$ that takes into account the scintillation efficiency and the quenching factor and relates the measured $\rm PE$ with the recoil energy of the scattered electrons $E_k$.
The ER acceptance and the conversion function in XENON1T are considered to be the same as for XENON100, both extracted from \cite{Aprile:2014eoa}. This is a conservative approach considering that XENON1T expects to increase the light collection in comparison with XENON100 \cite{Aprile:2015uzo}.
The differential number of events is then 
\begin{equation} \nonumber
\dfrac{dN_T}{dE_k}(m_S,|U_{Se}|^2)=\sum_t {\rm Acc(Conv}(E_k))n_t\dfrac{dN_t}{dE_k}(m_S,|U_{Se}|^2)\,,
\end{equation}
where $n_t$ is the number of electrons in the $t$ state. 
Any given values of the mass of the sterile neutrino $m_S$ and the mixing angle $|U_{Se}|^2$ generate a particular shape for the differential number of events in terms of the kinetic energy of the scattered electron $E_k$. This is shown in Fig.\ \ref{fig:dN} using $m_S=40$ keV and $|U_{Se}|^2=5\times10^{-4}$  for XENON1T; the background is also shown.

\begin{figure}[ht]
\centering
\includegraphics[width=0.475\textwidth]{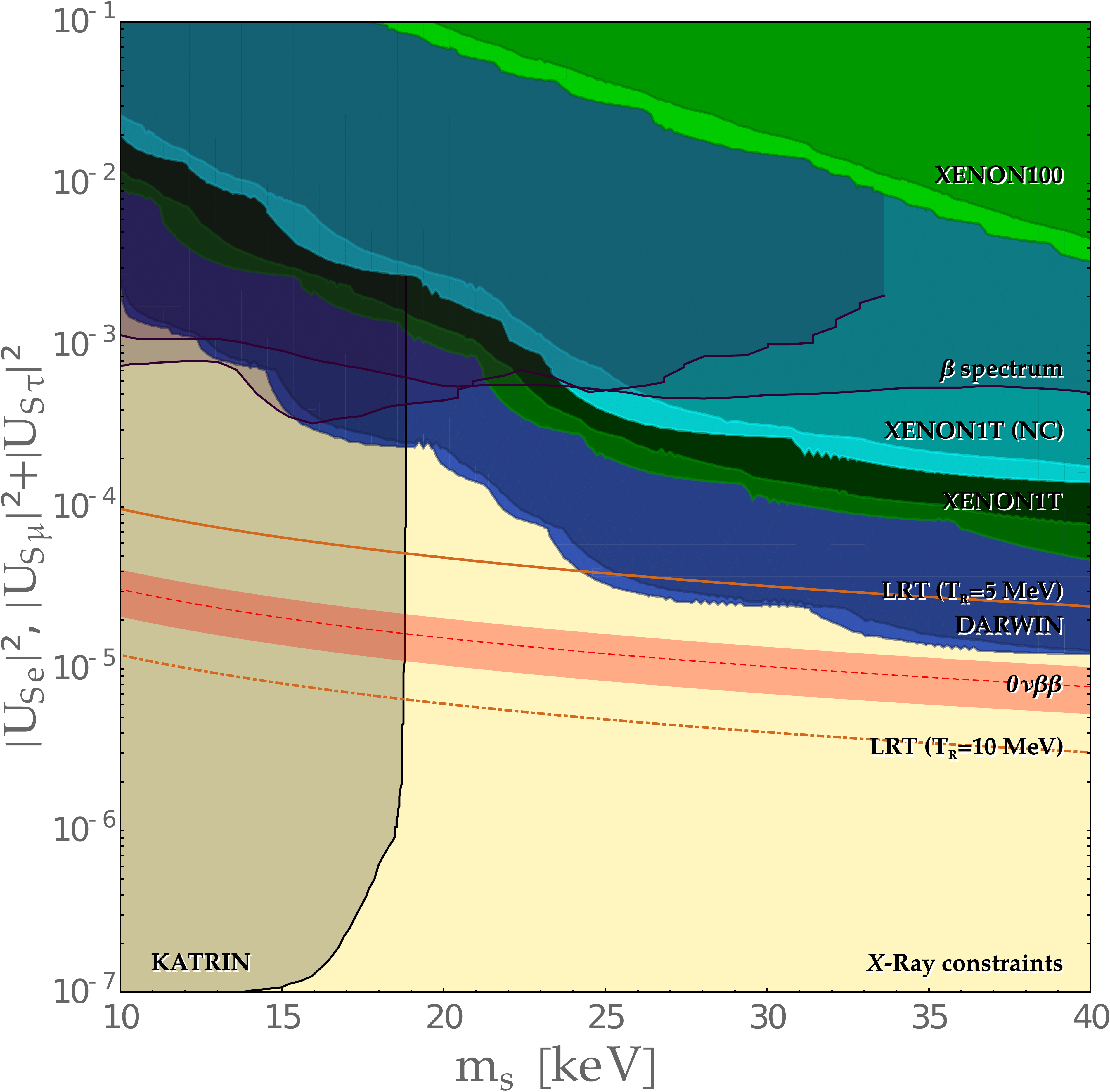}
\caption{Light Green: Sensitivity on sterile neutrino WDM parameters for XENON100 as a 
function of $m_S$ and $|U_{Se}|^2$. The contours delimit $90\%$ and $99.9\%$ C.L.\\
Dark Green: Equivalent for XENON1T;\\
Blue: Equivalent for DARWIN;\\
Purple: Current limits from analysis of $\beta$ spectrum of different radioisotopes 
\cite{Holzschuh:1999vy,Holzschuh:2000nj};\\
Black: Expected statistical sensitivity of a modified KATRIN setup (Fig.\ 11 in \cite{Mertens:2014nha});\\
Red dashed: Limits coming from $0\nu\beta\beta$ experiments \cite{Rodejohann:2011mu};\\
Orange solid (and dot-dashed): Excluded area for production in case of a low reheating temperature (LRT)
of $T_R=5$ MeV ($T_R=10$ MeV) \cite{Gelmini:2004ah};\\
Yellow: Constraint from $X$-ray searches \cite{Horiuchi:2013noa,Kusenko:2009up};\\
Turquoise denoted ``XENON1T (NC)'': limit on $|U_{S\mu}|^2 + |U_{S\tau}|^2$ if the sterile neutrino 
does not couple to electron neutrinos and only has neutral currents.}
\label{fig:Results}
\end{figure}

We restrict the analysis to the case in which the signal is entirely above the background and integrate to define
\begin{equation}
N_s := \int_{E_{\rm Th}}^{E_{0}} \dfrac{dN_T}{dE_k}dE_k
\quad\text{and}
\quad N_b := \int_{E_{\rm Th}}^{E_{0}} F_bdE_k\,.
\end{equation}
Here, $E_{\rm Th}$ is a lower energy threshold, 
which is set to 2 keV for XENON100 and 1 keV for XENON1T (this threshold is shown as a dashed black line in Figs.\ \ref{fig:ds} and \ref{fig:dN}). $E_0(m_S)$ is the point in which the signal $\frac{dN_T}{dE_k}$ intersects the background $F_b$, otherwise it is simply the upper 
bound of the electron energy allowed by kinematical constraints. 
For the DARWIN case, the background was extracted from \citep{Schumann:2015cpa} and extrapolated to the region of interest. The lower threshold and the acceptance function considered will be assumed 
as in the XENON1T case. 

From this block space analysis we define the significance in terms of a $\chi^2$-distribution, as a function of $N_s$ and $N_b$ 
\begin{equation}
\chi^2(m_S,|U_{Se}|^2):=\dfrac{(N_s(m_S,|U_{Se}|^2)-N_b(m_S,|U_{Se}|^2))^2}{N_b(m_S,|U_{Se}|^2)}.
\end{equation}
Imposing that $\chi^2\geq 4.60$ (13.82) for  $90\%$ ($99.9\%$) C.L.\ we obtain the region in terms of $m_S$ and $|U_{Se}|^2$ that can be excluded. 

Using the live-time shown in Eq.\ \eqref{MT} 
for XENON100 and XENON1T we obtain the light and dark green regions  
shown in Fig.\ \ref{fig:Results}. Notice that when compared to XENON1T, 
due to a larger background, higher threshold and a smaller value of $M \cdot T$, 
$|U_{Se}|^2$ must be $\sim\mathcal{O}(10^2)$ times larger to achieve the same effect in XENON100. 
In fact, for XENON100 the excluded area has been already excluded 
in other Earth-based experiments. However, XENON1T can set interesting limits on the parameters, 
and is much closer to detection than the capture of keV-scale WDM neutrinos in 
$\beta$-decaying nuclei \cite{Li:2010vy}. Additionally, using an exposure of 200 ton$\times$year for the DARWIN experiment we obtain the blue curve, which is a bit less than an order of magnitude better than the XENON1T case.
As a comparison, we show in the same figure the 
current exclusion limits for different 
Earth-based experiments in purple in which the $\beta$-spectrum of $^{63}$Ni \cite{Holzschuh:1999vy} and $^{35}$S \cite{Holzschuh:2000nj} was analyzed. 
The black area is the expected $90\%$ statistical exclusion limit of a differential 
measurement of 3 years with a modified setup of the 
KATRIN experiment \cite{Mertens:2014nha}. It 
would be several orders of magnitude more sensitive in the mixing angle, 
but the mass range is more limited. 
The red dashed line represents the limit using the null results from 
neutrinoless double beta decay ($0\nu\beta\beta$) experiments with the constraint 
$|U_{Se}|^2 \,m_S < (0.3 \pm 0.1)$ eV; see, 
for example, \cite{Rodejohann:2011mu}. This limit is almost one order of magnitude more sensitive in comparison to
XENON1T, but assumes that the 
neutrinos are Majorana particles (as in relic neutrino capture 
\cite{Long:2014zva} there will be a factor 2 smaller cross section for Dirac neutrinos) 
and that none of the many possible mechanisms for 
double beta decay interferes. 
For the shown mass range, the Dodelson-Widrow mechanism \cite{Dodelson:1993je} to produce the Dark Matter abundance sets a limit for $|U_{Se}|^2$ below $10^{-9}$, as derived in \cite{Asaka:2006nq}. This mechanism, in which the relic abundance is produced through the oscillation of active to sterile states in the early Universe is unavoidable, which means that if the $(m_S,|U_{Se}|^2)$ 
parameter space is in reach of the XENON1T or the DARWIN experiment, this would call for 
an alternative explanation for an otherwise overproduced relic
abundance (apart from having a too short lifetime). 
One possibility is, for example, to consider a low reheating temperature in inflationary models 
which would suppress the production of non-relativistic particles at 
$T\lesssim T_R$  \cite{Gelmini:2004ah}. The excluded area for $T_R=5$ MeV (and $T_R=10$ MeV) is 
shown in orange (and dot-dashed orange), and limits from XENON1T can be comparable. Another option is to assume the insertion of additional entropy in the system, as e.g.\ in Refs.\ \cite{Bezrukov:2009th,Nemevsek:2012cd}. 
The yellow area shows current astrophysical 
$X$-ray constraints \cite{Horiuchi:2013noa,Kusenko:2009up}. 
Finally, if the sterile neutrino does not couple to the electron neutrino, it can 
still have neutral current reactions. The (weaker) limit on 
$|U_{S\mu}|^2 + |U_{S\tau}|^2$ is also shown in the plot in turquoise. The limits from beta and 
double beta decay do not apply here. 

We can see that while astrophysical $X$-ray limits will remain the strongest bounds, 
the exclusion limit for the mixing angle with the electron neutrino 
obtained for current direct detection experiments can be up to one order of magnitude 
stronger in comparison with other Earth-based limits and up to two orders of magnitude using future experiments.

\section{Conclusions}
In this work, we have proposed a method to extract sensitive information on 
 sterile neutrino Warm Dark Matter from electron 
recoil data in direct detection experiments.
This has resulted in a way to constrain the mixing angle $|U_{Se}|^2$ and the sterile neutrino mass $m_S$, independently of cosmological exclusion limits and complementary to them. 
The constrained values of mass and mixing would in standard scenarios imply a too large dark 
matter density, and interesting and far-reaching non-standard scenarios would 
be needed in order to reconcile with observation. 
Considering two years of live-time for the XENON1T experiment, the mass range which is 
possible to exclude using the method analyzed is between 10 and 40 keV, while the limit 
on the square of the mixing angle can go down to $\sim 5\times10^{-5}$. On the other hand when 
the exposure is increased two orders of magnitude as in the case of the 
DARWIN experiment the mixing angle squared can be constrained down to $10^{-5}$.

\section*{Acknowledgments}
We thank Ludwig Rauch, Teresa Marrod\'an, Carlos Yaguna and Hiren Patel for 
helpful discussions. This work is supported by the IMPRS-PTFS (M. C.) 
and the DFG in the Heisenberg programme with Grant No. RO 2516/6-1 (W. R.). 

\vspace{-0.2cm}
\bibliography{DDDM}

\end{document}